\documentstyle[twocolumn,graphics,tighten,prl,aps]{revtex}
\newcommand{\kB}{k}
\newcommand{\kT}{\kB T}
\newcommand{\lB}{l_{\mathrm{B}}}
\newcommand{\nm}{\mathrm{nm}}

\newcommand{\dens}{\rho}
\newcommand{\subsalt}{{\mathrm{S}}}
\newcommand{\submacro}{{\mathrm{M}}}
\newcommand{\nc}{\dens_\submacro}
\newcommand{\ns}{\dens_\subsalt}
\newcommand{\eqref}[1]{(\ref{#1})}
\newcommand{\mysmall}[1]{{#1}}
\newcommand{\smallplus}{\mysmall+}
\newcommand{\smallminus}{\mysmall-}
\newcommand{\smallpm}{\mysmall\pm}
\newcommand{\np}{\dens_{\smallplus}}
\newcommand{\nn}{\dens_{\smallminus}}
\newcommand{\npm}{\dens_{\smallpm}}
\renewcommand{\ni}{\dens_{\mathrm{I}}}
\newcommand{\nz}{\dens_{\mathrm{Z}}}
\newcommand{\mydelta}{\delta}
\newcommand{\dnc}{\mydelta\nc}
\newcommand{\dnp}{\mydelta\np}
\newcommand{\dnn}{\mydelta\nn}
\newcommand{\dni}{\mydelta\ni}
\newcommand{\dpd}{\Delta\overline\psi}
\newcommand{\DH}{Debye-H\"uckel}

\newcommand{\ansatz}{\emph{ansatz}}
\renewcommand{\r}{{\mathbf r}}
\newcommand{\qstar}{q^{*}}
\newcommand{\figgap}{\vspace{12pt}}
\begin{document}
\draft
\wideabs{
\title{Charged colloids at low ionic strength: macro- or 
microphase separation?}
\author{Patrick B. Warren}
\address{Unilever Research Port Sunlight, Bebington, Wirral, 
CH63 3JW, UK.}
%
%
\maketitle
\begin{abstract}
Phase separation in charged systems may involve the replacement of
critical points by microphase separated states, or charge-density-wave
states. A density functional theory for highly charged colloids at low
ionic strength is developed to examine this possibility. It is found
that the lower critical solution point is most susceptible to
microphase separation. Moreover the tendency can be quantified, and
related to the importance of small ion entropy in suppressing phase
separation at low added salt. The theory also gives insights into the
colloid structure factor in these systems.
\end{abstract}
\pacs{PACS: 05.20.-y, 64.75.+g, 82.70.Dd}
} 

There has been much interest recently in statistical physics in charged
soft matter systems. Whilst much of this is biologically inspired (eg
DNA condensation~\cite{DNA}), there has also been a long standing
controversy in the colloid science community over the anomalous
behaviour of charge stabilised colloidal suspensions at low ionic
strengths \cite{cntrvrsy}. Recently \cite{vRH,PBW}, it has been
suggested that the anomalies in these systems may be understood in terms
of a miscibility gap which is the analogue of the vapour-liquid
coexistence in the restricted primitive model (RPM) of 1:1 electrolytes
\cite{RPM}. Arguably a theoretical consensus is emerging, although there
remain a number of competing theories \cite{compete}.

In all these examples, the crucial role of the counterions should not be
underestimated. For the case of charged colloidal suspensions, for
example, the theories show that the overall phase stability is almost
entirely due to the entropy of the counterions \cite{vRH,PBW}. The same
effect can be said to underpin the solubility of many water-soluble
polymers \cite{polsol}. The basic point is that bulk phase separation in
a charged system must be into electrically neutral phases. If this
involves significant fractionation of counterions, an entropic penalty
will be incurred which tends to suppress phase separation.

Clearly though, if bulk phase separation is suppressed, a possibility
still exists to undergo \emph{microphase separation}, where
electroneutrality can be broken locally. Critical points are
particularly susceptible to this, as first shown by Nabutovskii, Nemov
and Peisakhovich (NNP) using a Landau-Ginzburg theory \cite{NNP}.
Consider density fluctuations at wavevector $q$. At $q\to0$,
fluctuations are restricted to elecrically neutral combinations, but for
$q>0$ fluctuations can violate electroneutrality increasingly easily.
Thus one might expect some softening of the modes. Indeed, if the only
terms to $O(q^2)$ come from the long range Coulomb law, the analysis
below implies that all partial structure factors have a \emph{minimum}
at $q=0$. Since the $q=0$ partial structure factors diverge as one
approaches a critical point, this suggests there must exist regions
around critical points where a divergence at $q>0$ occurs first,
indicative that the critical behaviour is preempted by microphase
separation. However, there are often other terms arising at $O(q^2)$
from elsewhere which destroy the phenomenon. A closely analogous
microphase separation for polyelectrolytes in poor solvents has also
been examined \cite{polymicro}, but in the present study microphase
separation is driven purely by electrostatic effects.

Let us start by constructing a simplified but physically motivated model
for the anomalous behaviour in charged colloidal suspensions. Consider
the macroions as spheres of charge $Z$, diameter $2a$, and number
density $\nc$ (volume fraction $\phi=4\pi a^3\nc/3$). There are small
counterions and coions at number densities $\nn$ and $\np$ respectively.
The solvent is a dielectric continuum. Without loss of generality, I
suppose the small ions are univalent, and there is only one species of
counterion \cite{wlog}. Since the coions come from added salt, it will be
convenient to write $\np=\ns$. Overall, the system is electrically
neutral and $\nn=\ns+Z\nc$, but $\mydelta\npm$ will be retained for
fluctuations.

Each macroion polarises the surrounding electrolyte solution, and
becomes surrounded by a `double layer'. It has been shown by many
workers \cite{vRH,BSCM,others} that the self energy of the macroion with
its double layer, in \DH\ theory, is $(Z^2\lB\kT/2a)\times h(\kappa a)$.
In this $\lB=e^2/\epsilon\kT$ is the Bjerrum length, the function
$h(x)=1/(1+x)$ \cite{hnote}, and the Debye screening length,
$\kappa^{-1}$, is given by $\kappa^2 = 8\pi\lB\ni$ where $2\ni=\nn+\np =
Z\nc+2\ns$ is (twice) the ionic strength. This energy has a well known
interpretation: it corresponds exactly to a \emph{spherical capacitor},
charged $\pm Ze$, with one plate at the macroion surface and the second
a distance $\kappa^{-1}$ away \cite{capacitor}.

The simplest model free energy based on this is 
\begin{eqnarray}
&&{F}/{V\kT}=\ns\log\ns+(\ns+Z\nc)\log(\ns+Z\nc)\nonumber\\
&&\qquad\qquad\qquad+\nc\log\nc
+\nc\,(Z^2\lB/2a)\,h(\kappa a).\label{feq} 
\end{eqnarray} 
The first three terms are the ideal terms, and the last is the self
energy of macroions at number density $\nc$. The most important omission
from this is the contribution from the macroion-macroion interactions.
Whilst this plays a significant role in structuring the macroions, it
has been demonstrated elsewhere \cite{PBW} that it is \emph{less
significant} than the self energy, as regards the appearance of a
miscibility gap. Moreover, by leaving this contribution out of the
theory, we will see quite clearly how structure can develop in the
system in the absence of pair interactions.

\begin{figure}
\begin{center}
\figgap
\includegraphics{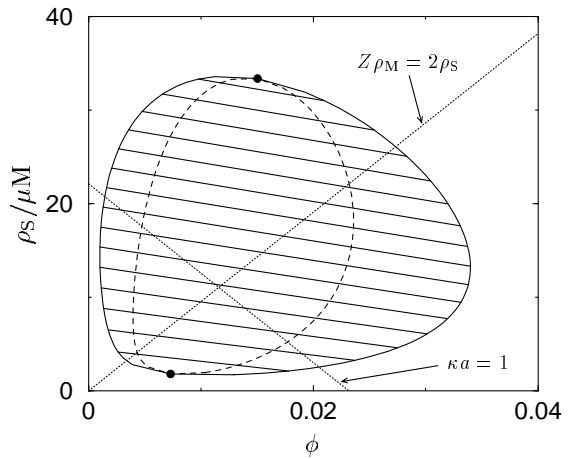}
\end{center}
\caption{Miscibility gap at low ionic strength for model free energy,
Eq.~\eqref{feq} in text, at $Z=10^3$ and $2a=100\,\nm$. Also marked are
the lines where the counterions and added salt contribute equally to the
ionic strength ($Z\nc=2\ns$), and where the macroion self energy
function varies most rapidly ($x=\kappa a=1$).\label{fig1}}
\end{figure}

A typical phase diagram corresponding to the above free energy is
shown in Fig.~\ref{fig1}, for $Z=10^3$ and $2a=100\,\nm$.  It
comprises a simple miscibility gap, limited above and below by
critical solution points as the salt concentration is varied.  The
gap occurs at very low ionic strengths, and only appears if the charge
on the macroions is sufficiently high ($Z\lB/a\agt13.4$ for 
$2a$ in the range 10--1000$\,\nm$).

Now, the NNP scenario could occur at either critical point. To examine
this therefore, I construct a density functional theory to correspond
to the free energy introduced above. The ideal terms become
$\kT\int\!d^3\r\,\dens_i(\r)\log \dens_i(\r)$ ($i=+,-,\submacro$), and
I introduce the \ansatz\ that the self energy generalises in the
obvious way to $\int\!d^3\r\,\nc(\r)\,f_N^{\mathrm{self}}(\r)$ where
$f_N^{\mathrm{self}}=(Z^2\lB\kT/2a)h(\kappa a)$ is the self energy per
particle evaluated using the local ionic strength at the particle
centre, $\kappa^2=8\pi\lB\ni(\r)$.
Finally, an electrostatic contrbution has to be added:
$\lB\kT\int\!d^3\r\,d^3\r'\,{\nz(\r)\nz(\r')}/{|\r-\r'|}$,
where $\nz(\r)=Z\nc(\r)+\np(\r)-\nn(\r)$ is the local charge density.

To examine the stability of the system against microphase separation,
expand the above density functional about the homogeneous state to
quadratic order.  For fluctuations at a wavevector $q$ this
results in
\begin{eqnarray}
&&\frac{\mydelta F}{V\kT}=\frac{|\dnp|^2}{2\ns}
+\frac{|\dnn|^2}{2(\ns+Z\nc)}
+\frac{|{\dnc}|^2}{2\nc}\nonumber\\
&&\qquad\qquad\quad+\frac{2\pi\lB}{q^2}|Z\dnc+\dnp-\dnn|^2\nonumber\\
&&\qquad\qquad\qquad+\frac{Z^2\lB}{2a}\Bigl[
8\pi^2\lB^2 a^4\nc h_1(\kappa a)|\dni|^2\label{dfeq}\\
&&\qquad\qquad\qquad\quad
-2\pi\lB a^2 h_2(\kappa a)(\dnc\dni^*+\dnc^*\dni)
\Bigl]\nonumber
\end{eqnarray}
The functions are $h_1(x) = (1+3x)/(x^3(1+x)^3)$ and $h_2(x) =
1/(x(1+x)^2)$. From this the macroion structure factor $S(q) =
\langle|\dnc(q)|^2\rangle$ is extracted in the standard way
\cite{slnote}. The behaviour of $S(q)$ is examined as a function of
$\phi$ and $\ns$, looking for the unstable regions in the
$(\phi,\ns)$-plane where $1/S(q)<0$.

At $q=0$ the spinodal instability region corresponding to the free
energy in Eq.~\eqref{feq} is recovered. For $q>0$ the region of
instability \emph{always expands}. This is because the $q$-dependence
arises solely from the long range electrostatic term, thus, as alluded
to above, $S(q)$ always has a minimum at $q=0$. But a clear difficulty
emerges when the behaviour for large $q$ is examined, since the
instability region expands to fill the entire plane; there is no
effective penalty against microphase separation at vanishingly small
wavelengths.

In fact the \ansatz\ for the self energy term has omitted an obvious but
crucial effect, namely one would not expect a macroion to be sensitive
to variations in the local ionic strength over distances much smaller
than its size. Motivated by weighted local density theories for liquids
\cite{wlda}, I therefore introduce an additional \emph{smoothing}
\ansatz. It turns out that the precise form does not matter greatly, for
instance one can replace the local ionic strength by $\overline\ni(\r) =
\int\!d^3\r'\, w(|\r-\r'|)\, \ni(\r')$ where $w(r)$ is a smoothing
kernel of range $a$ \cite{anote}, but equally one might smooth $\kappa$
or $f_N^{\mathrm{self}}$. All forms result in the appearance of extra
multiplicative factors, $w_1(qa)$ and $w_2(qa)$, in the last two terms
of Eq.~\eqref{dfeq}. The $w_i(qa)$ are related to the Fourier transform
of $w(r)$, and satisfy $w_i\to1$ at $q\to0$, $w_i\to0$ at $q\to\infty$.
(More generally I expect $h_i(\kappa a,qa)$ such that $h_i\to0$ for
$qa\to\infty$.) With this \ansatz, progress can be made without
developing a detailed theory by investigating various possibilities for
$w_i$. The results reported below have been carried out assuming
$w_1=w_2=\exp(-\alpha q^2a^2)$ with $\alpha$ a numerical prefactor of
order unity. Very similar results are obtained for $w_i=1/(1+\alpha
q^2a^2)$.

\begin{figure}
\begin{center}
\figgap
\includegraphics{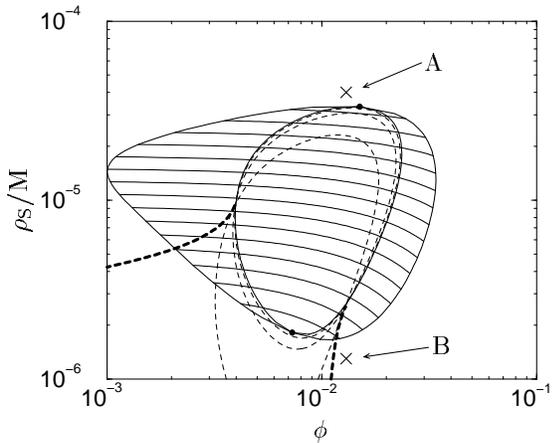}
\end{center}
\caption{Phase diagram in double logarithmic plot, showing as dashed
lines the contours of divergence of $S(q)$, from $q=0$ (identical to
original spinodal), through $qa=0.05$, 0.1, 0.2 (light dashed lines) to
$q=\qstar$ (heavy dashed line) where the maximum in $S(q)$ diverges
(typically $\qstar a\protect\alt0.5$). The structure factors
corresponding to the marked points are given in
Fig.~\ref{fig3}.\label{fig2}}
\end{figure}

\begin{figure}
\begin{center}
\figgap
\includegraphics{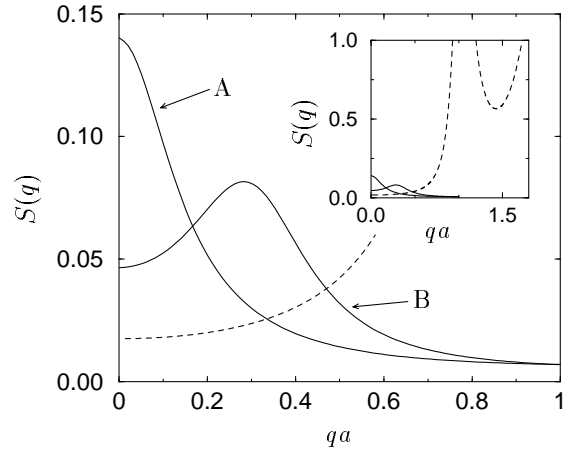}
\end{center}
\caption{Structure factor in vicinity of miscibility gap. The state
points are those marked in Fig.~\ref{fig2}. The dashed line
is the structure factor expected from the interparticle repulsions at a
scaled volume fraction around freezing (all normalised such that
$S(q)\to1$ at $q\to\infty$).\label{fig3}}
\end{figure}

Typical results from this modified theory are shown in
Figs.~\ref{fig2} and \ref{fig3}, for $\alpha=1$. In Fig.~\ref{fig2},
the spinodal instability at $q=0$ is recovered as before. For $q>0$, the
instability region is again expanded in the vicinity of the lower
critical point, but is now \emph{reduced} in the vicinity of the upper
critical point. For $qa\gg1$ the instability disappears completely,
since the self energy which drives the instability is now insensitive to
short wavelength fluctuations. Below the lower critical point,
therefore, there is a region (delimited by the heavy dashed line in
Fig.~\ref{fig2}) where $S(q)$ diverges at some $\qstar>0$,
corresponding to the appearance of microphases.

These results are reflected in the macroion structure factors, two
examples of which are shown in Fig.~\ref{fig3}. Near the upper critical
point the structure factor turns up to a maximum as $q\to0$, developing
a divergence at $q=0$ as the (mean field) critical point is approached.
Near the lower critical point a peak appears in $S(q)$ at $\qstar>0$.
The peak diverges as one approaches the boundary of the microphase
instability region. Note that the appearance of a peak in the macroion
structure factor $S(q)$ is unusual because there are no direct macroion
interactions in the theory as constituted above. This shows how
structure can arise in a charged system independent of the existence of
(effective) pair interactions.

To relate the results to possible experiments, one should of course
investigate the significant contribution to the structure factor from
the omitted macroion interactions. Elsewhere it is argued that, at these
low ionic strengths, the macroions are effectively a one-component
plasma (OCP) in the strong coupling limit \cite{PBW}, whose structure
factor resembles that of hard sphere (HS) fluid close to the freezing
transition \cite{HSfreeze}. A rescaled $S(q)$ for HS at freezing is
compared with the present calculations in the inset in Fig.~\ref{fig3}.
The additional structure arising from the self energy theory above
appears to lie well within the first peak, and moreover the amplitudes
match quite closely. Thus the appearance of a maximum at $q=0$ or a peak
at very low $q$ may well be experimentally observable. This may be the
explanation of the anomalously large $S(q\to0)$ reported recently for
colloidal suspensions at low ionic strengths \cite{expts}.

Let us turn to the effect of $\alpha$. Recall that $\alpha$ is a measure
of the degree of smoothing: the range of the smoothing kernel is $\sim
a\sqrt{\alpha}$. All the results discussed above were at $\alpha=1$. If
$\alpha\alt0.338$, microphases appear at the \emph{upper} critical point
too. On the other hand, if $\alpha\agt1.961$ microphase separation at
the lower critical point disappears. One would expect increasing
$\alpha$ to suppress microphase separation, since greater smoothing is
bound to reduce $\qstar$, but the difference between the two critical
points is suggestive. These critical values of $\alpha$ are more general
than the assumed form of $w_i(q)$, since they only depend on the $q^2$
coefficient in the expansion $w_i(q)=1-\alpha q^2 a^2 + O(q^4)$
\cite{alphnote}. They are a measure of the susceptibility of the
critical point to replacement by microphases (if $\alpha$ could be
treated as a control variable, the critical values would correspond to
Lifshitz points in the phase diagram).

This calculation sheds light on the reason for a \emph{closed loop}
miscibility gap. As $\ns$ is increased, an upper critical point is
expected since the self energy ceases to be strongly state point
dependent for $\kappa a\gg1$ or $\ns\gg Z\nc$, where it simply shifts
the macroion chemical potential (see lines in Fig.~\ref{fig1}).  As
$\ns\to0$ though, it is perhaps unexpected to encounter a
\emph{second} critical point. Its appearance appears to be connected
to the small ion entropy effect discussed in the introduction. The
evidence for this is twofold. Firstly, as already commented upon, the
lower critical point is more susceptible to microphase
separation. This is in accord with the idea that phase separation is
suppressed by small ion entropy, since in microphase separation, the
system gains entropy by distributing the small ions more uniformly
than would be allowed if strict electroneutrality had to be satisfied
at each point.

The second piece of evidence concerns the rate at which the \emph{Donnan
potential difference} $\dpd$ vanishes as one approaches the critical
point. Recall that $\dpd$ arises because the interface can acquire a
dipole moment density (per unit area). In the present system, a dipole
moment density appears to arise because the jump in small ion densities
is spread out more broadly than the jump in macroion densities (although
this remains to be confirmed with a detailed calculation
\cite{dpdnote}). Remarkably, one can calculate $\dpd$ without detailed
knowledge of the structure of the interface \cite{PBW}. I find that
$\dpd$ vanishes as $\Delta\phi/\phi_{\mathrm{crit}}$ as the critical
points are approached, with a constant of proportionality $\approx8.00$
for the upper critical solution point, and $\approx16.9$ for the lower
one. This again indicates the growing importance of small ion entropy
(broadening the jump in small ion densities) as the lower critical point
is approached. 

Note that $\dpd$ is an order parameter which strictly vanishes in
symmetric models such as the RPM. Apart from general remarks by
Nabutovskii and Nemov \cite{NN}, the critical behaviour of asymmetric
primitive models seems to have received much less attention than the
RPM \cite{RPM}, and there may be interesting effects connected to a
non-vanishing $\dpd$.

I thank C. Holmes and M. E. Cates for useful discussions, and P.
Schurtenberger for sending data prior to publication.


\begin{references}
\bibitem{anote}{Note that in the phase separation region there is
only one significant length scale since $\kappa^{-1}\sim a$.}
\bibitem{alphnote}{For simplicity only the case $\alpha_1=\alpha_2$ is
treated; in general one expects a critical line in the
$(\alpha_1,\alpha_2)$ plane.}
\bibitem{dpdnote}{Obviously, the density functional theory proposed here
provides a natural starting point for such a calculation, which will be
the subject of future work.}
\end{references}
\end{document}